%% file: mcmc_p2.tex
\documentclass[twocolumn]{aastex631}

\usepackage{amsmath} 
\usepackage{graphicx} 
\usepackage{textcomp}
\usepackage{gensymb}
\usepackage{natbib} 
\usepackage{longtable}
\usepackage{multirow}
\usepackage{float}
\usepackage{enumitem}

\begin{document}
\def\teff{$T_{eff}$}
\def\cs{$\chi^{2}$}
\def\rsun{$R_{\odot}$}
\def\msun{$M_{\odot}$}
\def\rstar{$R_{\star}$}
\def\rearth{$R_{\earth}$}
\def\av{$A_{V}$}
\def\kep{\textit{Kepler}}
\def\emcee{\texttt{emcee}}
\newcommand\val[3]{$#1^{+#2}_{-#3}$}
\def\ee{$\eta_{\earth}$}

\graphicspath{{figures/}}
\shorttitle{Binary Planet Hosts II.}
\shortauthors{Sullivan \& Kraus}

\title{Revising Properties of Planet-host Binary Systems II: Apparent Near-Earth Analog Planets in Binaries Are Often Sub-Neptunes \footnote{Based on observations obtained with the Hobby-Eberly Telescope, which is a joint project of the University of Texas at Austin, the Pennsylvania State University, Ludwig-Maximilians-Universität München, and Georg-August-Universität Göttingen.}}

\author[0000-0001-6873-8501]{Kendall Sullivan}
\altaffiliation{NSF Graduate Research Fellow}
\affil{University of Texas at Austin, Austin, TX 78712, USA}

\author{Adam L. Kraus}
\affil{University of Texas at Austin, Austin, TX 78712, USA}

\correspondingauthor{Kendall Sullivan}
\email{kendallsullivan@utexas.edu}

\begin{abstract}
Identifying rocky planets in or near the habitable zones of their stars (near-Earth analogs) is one of the key motivations of many past and present planet-search missions. The census of near-Earth analogs is important because it informs calculations of the occurrence rate of Earth-like planets, which in turn feed into calculations of the yield of future missions to directly image other Earths. Only a small number of potential near-Earth analogs have been identified, meaning that each planet should be vetted carefully and then incorporated into the occurrence rate calculation. A number of putative near-Earth analogs have been identified within binary star systems. However, stellar multiplicity can bias measured planetary properties, meaning that apparent near-Earth analogs in close binaries may have different radii or instellations than initially measured. We simultaneously fit unresolved optical spectroscopy, optical speckle and near-infrared AO contrasts, and unresolved photometry, and retrieved revised stellar temperatures and radii for a sample of 11 binary \textit{Kepler} targets that host at least one near-Earth analog planet, for a total of 17 planet candidates. We found that 10 of the 17 planets in our sample had radii that fell in or above the radius gap, suggesting that they are not rocky planets. Only 2 planets retained super-Earth radii and stayed in the habitable zone, making them good candidates for inclusion in rocky planet occurrence rate calculations.
\end{abstract}

\keywords{}
\accepted{August 12 2022}

\section{Introduction}\label{sec:intro}
As the number of known exoplanets has grown, increasingly complex analyses of the sample of exoplanets have become possible. However, measuring the basic demographics of exoplanets remains a challenging topic, even with more than 5000 confirmed exoplanets \citep{exoplanetarchive}, because each planet search has its own biases and internal systematics impacting target selection and sensitivity to different regions of parameter space. One of the key demographic properties of exoplanets is the occurrence rate of Earth-like (i.e., small and rocky) planets in the habitable zone (HZ) of their host star, often referred to as \ee. The HZ is defined as being the region around the star where liquid water would be possible on the surface of a rocky planet \citep{Huang1959, Hart1978, Kasting1993}. 

Calculating \ee\ requires measurements of the radii and masses of planets and their host stars to constrain the composition and size of the planets and the location of the star's HZ. These tasks are difficult in their own right, but \ee\ is even more difficult to measure directly because detection of Earth analog systems is at the edge of current observational capabilities. Estimates of \ee\ have varied by more than an order of magnitude over the last decade \citep[e.g.,][]{Petigura2013, Foreman-Mackey2014, Silburt2015, Kaltenegger2017, Zink2019, Bryson2021}, and are not statistically robust, because the small number of observed Earth analog systems leads to large uncertainties in \ee\ completeness corrections. Therefore, any changes to the small sample of Earth analog planets will have implications for the calculation of \ee.

More robust calculations of \ee\ will be possible with a larger sample of near-Earth analog planets, but these calculations also require careful vetting of the current candidates to prevent over- or under-estimates of \ee. This vetting is particularly important as the science specifications for future Earth-like planet direct imaging missions are being developed, because one key goal of those missions will be to image Earth-like planets and characterize their atmospheres. It is necessary to have an accurate calculation of \ee\ to determine the survey size and expected yield of missions like the proposed 6m space telescope recommended by the 2020 Astronomy Decadal Survey.

One under-examined population of potential near-Earth analog planets are those in binary star systems. Binaries impact planets and planet characterization in a variety of ways. Close binary systems ($\rho < 50$ au) suppress planet occurrence rate \citep{Kraus2016, Moe2021}, and multiplicity can bias inferred stellar (and therefore planet) properties \citep{Furlan2020, Sullivan2022b}. Many past studies of planet demographics that are not explicitly focused on multiplicity have attempted to remove binaries from their sample, but systems are often found to be multiples after the fact \citep[e.g.,][]{Furlan2017}. Although planet occurrence is suppressed in binaries, $\sim$ 50\% of solar-type stars are in binary systems \citep{Duquennoy1991, Raghavan2010}, so by neglecting binary-star planet hosts, a substantial number of planets are ignored. Planets in binary systems are also relatively rare, so ignoring them also entails neglecting a potentially interesting population that evolved and survived in complex and extreme environments.

Alternatively, some planets in binaries were included in \ee\ calculations using properties that were measured without fully accounting for the effects of the binary. For example, an incorrect temperature measurement for the planet host can alter the location of the stellar HZ and change the inferred stellar radius, changing the planet's instellation flux and the inferred planetary radius. An undetected binary can have a temperature that is measured to be $\sim$200-300 K cooler than the true temperature of the primary star, and several hundred K hotter than the secondary temperature \citep[e.g.,][]{Furlan2020, Sullivan2022b}. Similarly, luminosity assumptions propagate to inferred stellar radii, and thus to inferred planetary radii. Both of these changes impact the inclusion of a given planet in the \ee\ calculation. 

Some studies \citep[e.g.,][]{Law2014, Kraus2016, Furlan2017, Ziegler2018} have used high-resolution imaging to identify companions to \kep\ host stars, but these groups did not perform spectroscopy to characterize the companions they found. Conversely, the California Kepler Survey \citep{Petigura2017} performed high-resolution spectroscopy on a subsample of \kep\ targets, but intentionally avoided known binaries. A few systems of close binaries and planets have been studied in depth \citep[e.g.,][]{Cartier2015, Barclay2015}, but those analyses have typically been specialized to the quirks of individual data sets and restricted to particularly interesting systems. In general, obtaining the observations needed to adequately characterize both components of a planet-hosting binary (extensive high-resolution imaging or echelle spectroscopy, to spatially or spectrally distinguish the components) is observationally expensive and limited to only the brightest targets. Therefore, no study has intentionally spectroscopically observed \kep\ binary planet hosts.

We have developed a Markov Chain Monte Carlo (MCMC) technique to determine the temperatures, radii, and luminosities of unresolved binary stars via SED fitting of unresolved low-resolution spectroscopy, unresolved photometry, and pre-existing high-resolution imaging measurements \citep{Sullivan2022b}. Low-resolution spectroscopy is less expensive than high-resolution observations, and contrasts are typically already available for previously-validated planetary candidates, so our technique greatly increases the number of binary systems that can be accurately deconvolved, and thus have their planets characterized. We have used the Low Resolution Spectrograph (LRS2) on the Hobby-Eberly Telescope (HET) at McDonald Observatory to observe 11 \kep\ Objects of Interest (KOIs) that host potential Earth analogs, then used this new fitting technique to more accurately determine the system characteristics and revise the inclusion of these systems in the \ee\ calculation. 

\section{Sample Selection}\label{sec:data}
We selected our sample of Earth analog planets in binary systems using system characteristics from \citet{exofop} \footnote{\url{exofop.ipac.caltech.edu/}}. We queried the database for systems where at least one planet had a radius of $R<1.8$R$_{\earth}$ and an instellation flux $S < 5 S_{\earth}$, and restricted the search to only include classifications of ``planet candidate'' or ``confirmed planet''. We cross-matched that list of near-Earth analogs with the high-resolution imaging compilation of \citet{Furlan2017}, which compiled observations of many KOIs using both original and literature high-resolution imaging observations. 

We restricted our cross-matched sample to binary systems with separation $\rho < 2 \arcsec$, possessing more than one measured contrast, and with contrast $\Delta$mag $< 3.5$ mag in at least one band. These choices ensured that both binary components would be included in the single unresolved spectrum, and that there would be adequate flux from the secondary present in the spectrum to accurately measure its properties. Finally, we visually inspected any available high-contrast imaging data for sources to remove any systems that appeared to have a false positive binary identification (did not show any apparent secondary companion), but did not identify any false positives. 

We identified three systems (KOI-3284, KOI-3456, and KOI-3497) that were classified as binaries but appeared to be triple systems, and we excluded those systems from the analysis because we did not have contrasts for the tertiary component. The triple system KOI-2626 did have contrasts for the tertiary, but it had optical contrasts from both HST and Gemini in similar filters that were not consistent. Because it was not clear which data set should be adopted, we chose to remove KOI-2626 from our sample. We were left with a sample of 11 KOIs hosting Earth analog planets in multi-star systems.

\begin{deluxetable*}{CCCCCCCCCC}
\tablecaption{Target list, separations, and existing direct imaging contrasts}
\tablecolumns{9}
\tablewidth{0pt}
\tablehead{
\colhead{KOI} &
\colhead{Sep. (\arcsec)} & \colhead{$\Delta m_{LP600}$} & \colhead{$\Delta m_{562 nm}$} & \colhead{$\Delta m_{692 nm}$}
& \colhead{$\Delta m_{880 nm}$} & \colhead{$\Delta m_{J}$} & \colhead{$\Delta m_{K}$} & \colhead{$\Delta m_{\text{Br}\gamma}$}
}
\startdata
1422\tablenotemark{a} & 0.22 & \nodata & \nodata & 1.720 \pm 0.150 & 1.620 \pm 0.150 & 1.078 \pm 0.038 & 1.163 \pm 0.021 & \nodata\\
2124\tablenotemark{a} & 0.06 & \nodata & 0.510 \pm 0.15 & \nodata & 0.180 \pm 0.150 & \nodata & 0.011 \pm 0.010 & \nodata\\
2298 & 1.52 & 2.08$\pm$0.14 & \nodata & \nodata & \nodata & \nodata & \nodata & 1.3$\pm$0.02\\
2418 & 0.11 & \nodata & \nodata & 3.22 $\pm$ 0.15 & 2.94 $\pm$ 0.15 & \nodata & 2.509 $\pm$ 0.062 & \nodata\\
2862\tablenotemark{a} & 0.63 & 0.17 \pm 0.05 & \nodata & \nodata & \nodata & \nodata & -0.001 \pm 0.010 & \nodata\\
3010\tablenotemark{a} & 0.33 & \nodata & \nodata & 0.74 \pm 0.15 & 0.01 \pm 0.15 & \nodata & 0.245 \pm 0.052 & \nodata\\
3255 & 0.18 & \nodata & \nodata & 0.520 \pm 0.150 & 0.400 \pm 0.150 & \nodata & 0.115 \pm 0.012 & \nodata\\
4986 & 0.17 & \nodata & \nodata & \nodata & \nodata & 1.703 \pm 0.046 & \nodata & 1.413 \pm 0.010\\
5545 & 0.08 &\nodata & 0.620 \pm 0.150 & \nodata & 0.500 \pm 0.150 & 0.260 \pm 0.031 & \nodata & 0.383 \pm 0.25\\
5971 & 0.04 & \nodata & 0.00 $\pm$ 0.15 & \nodata & 0.72 $\pm$0.15 &  \nodata & \nodata & \nodata\\
7235 & 0.11 & \nodata & \nodata & \nodata & \nodata & 0.019 \pm 0.032 & \nodata & 0.082\pm 0.034\\
\enddata
\tablenotetext{a}{Included in \citet{Sullivan2022b}.}
\tablecomments{Contrasts taken from \citet{Horch2012}, \citet{Dressing2014}, \citet{Everett2015}, \citet{Baranec2016}, \citet{Kraus2016}, and \citet{Furlan2017}.}
\label{tab:targets}
\end{deluxetable*}

Table \ref{tab:targets} lists the separation and contrasts for each system from the ExoFOP. \citet{Furlan2017} took their own observations as well as compiling results from a number of different works. Among these other surveys, \citet{Horch2012} used the DSSI instrument on Gemini North; \citet{Dressing2014} used the ARIES instrument on the MMT; \citet{Everett2015} used DSSI speckle imaging at Gemini North and NIR AO imaging at several different sites; \citet{Kraus2016} used NIRC2 at Keck Observatory; and \citet{Baranec2016} used Robo-AO on the Palomar Observatory 1.5m telescope.

We removed the $\Delta m_{[880]}$ contrast for KOI-3010 from our analysis because it was inconsistent with other contrasts in the literature, such as the Hubble Space Telescope F775W contrast presented in \citet{Gilliland2015} ($\Delta m_{[775W]}$ = $0.294 \pm 0.05$ mag, whereas $\Delta m_{[880]}$ = $0.01 \pm$0.15 mag). However, the $\Delta m_{[880]}$ contrast for KOI-3010 is included in Table \ref{tab:targets} for completeness. The $\Delta m_{J}$ value for KOI-4986 presented on the ExoFOP seemed inconsistent with the $\Delta m_{K}$ (assuming both components are main sequence stars) so for that system we reanalyzed the Keck AO images following \citet{Kraus2016} and used the resulting contrasts, which are the values listed in the table.

\section{Observations using LRS2 on HET} \label{sec:obs}
We observed all systems using the red setting (6500 $< \lambda < $ 10500 \AA) of the second-generation low-resolution spectrograph (LRS2-R; \citealt{Chonis2014, Chonis2016}) at the Hobby-Eberly Telescope (HET) at McDonald Observatory. The observations were taking in queue observing mode between 20210402 and 20210920. We calculated integration times using the facility integration time calculator, choosing times that were either comparable to the typical overhead (300s) or long enough to achieve a predicted signal-to-noise ratio (SNR) of 100 at 7500\AA\ (typically 450-600s), whichever was longer. Because our observations did not require resolved sources, we set a high seeing threshold of 2.5$\arcsec$, which was sufficient to obtain high-quality spectra for our bright targets.

LRS2 is an integral field spectrograph with a 12\arcsec$\times$ 6\arcsec\ field of view continuously tiled by 0.6\arcsec\ hexagonal lenslets. It has two possible observing modes, LRS2-B and LRS2-R, corresponding to pairs of observations in the blue or red ends of the optical spectrum. Either setting of LRS2 observes in two arms simultaneously: UV and orange for LRS2-B and red and far-red for LRS2-R. Because the majority of our sources were relatively cool stars, and therefore had spectra that peaked in the red or NIR, we choose to observe in the LRS2-R setting. 

The spectra from LRS2-R were reduced using the HET LRS2 pipeline, \texttt{panacea} (G. Zeimann et al. 2022, in preparation), which was possible because the binary sources had small enough separations to appear unresolved in the LRS2 data cubes. The primary steps in the reduction process are bias subtraction, dark subtraction, fiber tracing, fiber wavelength evaluation, fiber extraction, fiber-to-fiber normalization, source extraction, and flux calibration. Differential atmospheric refraction is corrected at each wavelength. Although LRS2-R observes in two arms, red (6500 $< \lambda < $ 8470 \AA) and far-red (8230 $< \lambda < $ 10500 \AA), the telluric contamination was severe in the far-red arm and the SNR was low because we optimized our observations for the red arm. Therefore, we only used the red arm for our analysis.

The source is modeled using a two-dimensional Gaussian profile fit to a synthetic image at the highest signal to noise in the spectrum. The profile for the source is used for an optimal weighted extraction \citep{Horne1986} clipped at an aperture of 2.5 times the seeing.  We truncated the data by 50 \AA\ on the red end of the spectrum to compensate for a reduction in throughput caused by shifts in the pivot wavelength of the dichroic separating the red and far-red arms of the instrument. When comparing to models, we normalized with a low-order polynomial to avoid potential uncertainties if the continuum slope of the spectrum was affected by instrumental errors.

\section{Analysis Methods}
\subsection{Telluric Removal} 
We did not observe telluric standard stars during our observing campaign, which meant that we needed to correct for atmospheric absorption in the data using an alternative method before retrieving the properties of the binaries in our sample. To correct for telluric absorption, we used Earth atmosphere models to perform a first-order correction to the data, then used a combination of error weighting and spectral masking to compensate for inaccuracies in the telluric models.

The Earth's atmospheric absorption in the LRS2-R wavelength range is dominated by water and O$_{2}$ bands, most of which are present in the spectrum as deep and broad features. To perform an initial fit to the telluric features in the data, we generated a grid of telluric models with humidity levels ranging from 5\% to 95\% in increments of 10\% using the Earth atmospheric modeler implemented in \texttt{TelFit} \citep{Gullikson2014} and using atmospheric conditions from a typical observation. We were able to generate a single grid of telluric models for all observations because the HET has a fixed altitude of $55\degree$, meaning that all HET observations are taken at approximately the same airmass of $\sec{(z)} \sim 1.22$. Because the airmass is nearly constant, we were able to hold the oxygen abundance of the models constant and only vary the humidity. This method was sufficient for all our data, but introducing another fit axis to compensate for changing oxygen column depth would have been possible if our approximation was not successful. 

Using the grid of telluric models at different humidities, we performed a least-squares fit to the data using the \texttt{L-BFGS-B} algorithm implemented in \texttt{scipy.optimize.minimize}. At each fitting stage, we calculated the reduced-\cs\ between the data and a model that was a composite of a telluric spectrum created using the telluric grid and a single stellar spectrum created using the BT-Settl stellar atmosphere models \citep{Allard2013, Rajpurohit2013, Allard2014, Baraffe2015} with the \citet{Caffau2011} line list. We allowed the fitting algorithm to generate telluric models at any humidity within the physical range of [5\%, 95\%], and set our initial humidity guess at 50\%. We set the initial temperature guess to be the ExoFOP temperature, which was measured as part of the Kepler Input Catalog \citep{Brown2011}, and allowed the fitter to explore a range of 500 K on either side of the input temperature. We fixed the surface gravity at a value of $\log{g} = 4.75$, which is appropriate for main-sequence dwarfs (e.g., 5 Gyr MESA Isochrones and Stellar Tracks (MIST) stellar evolutionary models; \citealt{Paxton2011, Paxton2013, Paxton2015, Dotter2016, Choi2016}). 

\begin{figure}
    \plotone{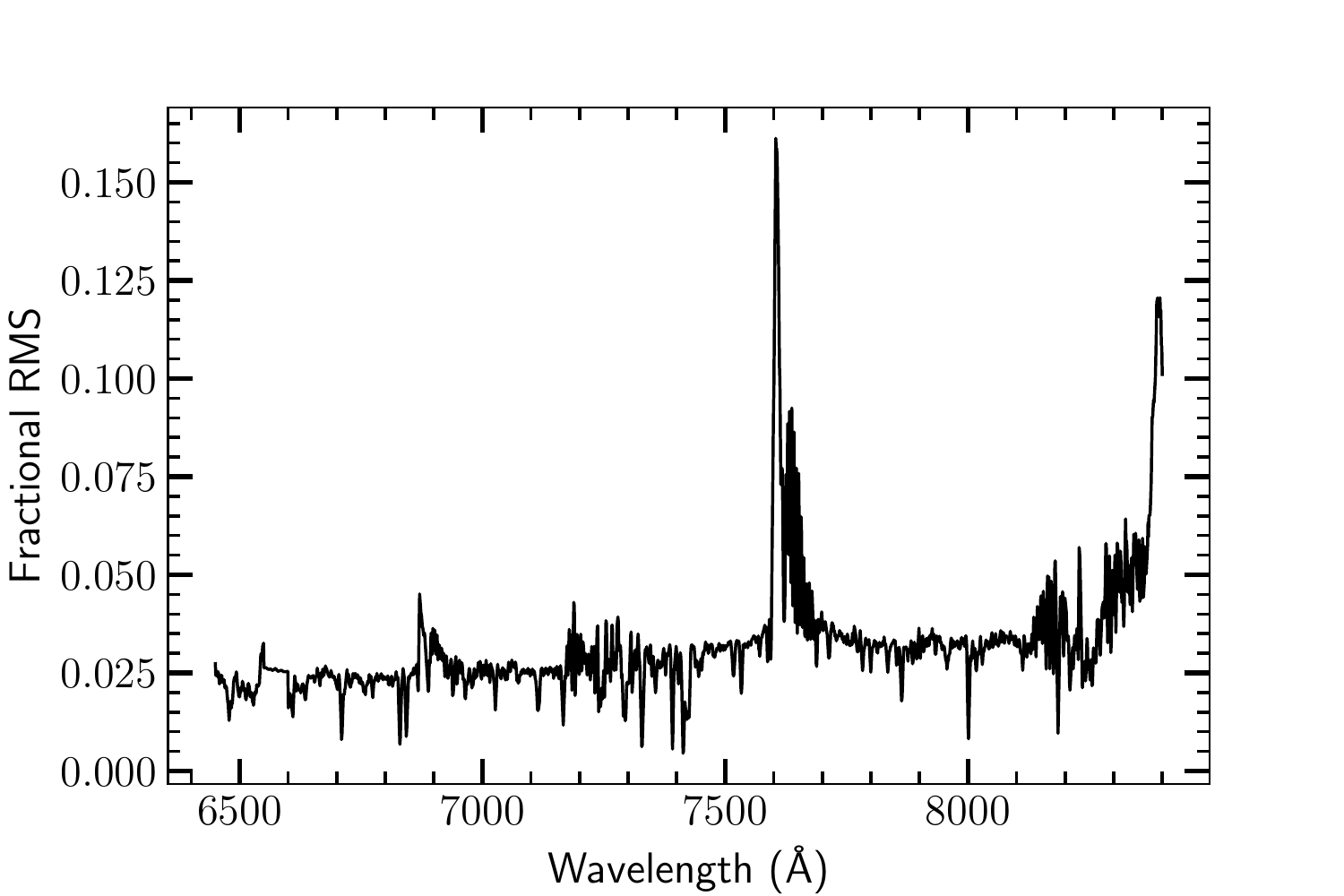}
    \caption{The RMS of the residuals from a sample of three telluric-corrected A0 stars. The RMS is typically less than 5\%, indicating that our method of telluric correction using models typically only introduces a small error increase into the spectrum.}
    \label{fig:a0 rms}
\end{figure}

After finding a best-fit humidity value, we divided the stellar spectrum by the best-fit telluric spectrum. The telluric models are imperfect, so this correction introduced additional error into the spectra. To assess the degree of error introduced by the telluric correction, we performed the same fitting method as described above on three telluric standard stars, which were the only targets available in the HET/LRS2 archive. Because telluric standard stars (typically A0 spectral type) have very few stellar spectral lines, virtually all of the residual after the telluric correction and stellar spectral fit was noise introduced by the correction process, rather than being a combination of imperfectly fit stellar and telluric lines. We assumed that the standard deviation of the sample at each pixel was representative of the noise introduced at that pixel even in a best-case correction, and so inflated the error of the observed data by the standard deviation of the residuals left after telluric-correcting the standard star sample. These increases were typically on the order of 2.5\%, and rarely more than 5\%. Figure \ref{fig:a0 rms} shows the weighting with which we modified the error on the observed science spectra, and the uncertainties in the figure also implicitly capture any target-to-target uncertainties in the flux normalization of LRS2-R. 

For spectral regions where the noise introduced by the telluric correction was larger than $\sim$5\%, we imposed a mask on the data during the final fitting process. We masked the following spectral regions: $6860-6880$\AA, $7600-7660$\AA, and $8210-8240$\AA. The first two regions contain strong oxygen absorption bands that are poorly modeled and difficult to properly correct, while the third region is a strong line in a water feature. In plots of the data, these regions are marked with gray bands to indicate that they were not included in the two-component fitting process.

We could have implemented \texttt{TelFit} \citep{Gullikson2014} or another telluric-fitting code simultaneously with our two-component fitting method, but retrieving the atmospheric model in \texttt{TelFit} is relatively slow and so would have significantly slowed down the two-component fitting process. We found that in lieu of that slightly more robust technique, this method produced reliable and consistent results for systems with HET observations that we had previously characterized in \citet{Sullivan2022b}, so we assumed it was an adequately rigorous level of telluric correction for our full HET sample. 

\subsection{Two-component Fitting Method}
The majority of our analysis method, alongside detailed initial validation tests, is described in detail in \citet{Sullivan2022b}, but is briefly summarized here for completeness. In contrast to the analysis in \citet{Sullivan2022b}, we included a search for the best-fit extinction value for each system, since some of the systems were at large distances or were close to the galactic plane and so had non-negligible extinction \citep{Green2019}.

We assembled a three-component data set for each system: a moderate-resolution composite spectrum of the system; unresolved photometry collected from 2MASS ($JHK_{s}$; \citealt{Skrutskie2006}) and the \kep\ Input Catalog (KIC $r'i'z'$; \citealt{Brown2011}); and contrasts collected on the ExoFOP website from various sources described in Section \ref{sec:data}. We fit the combined data set using the BT-Settl stellar atmosphere models \citep{Allard2013, Rajpurohit2013, Allard2014, Baraffe2015} with the \citet{Caffau2011} line list\footnote{\url{https://phoenix.ens-lyon.fr/Grids/BT-Settl/CIFIST2011/}}. 

Using the model spectra, we found the best-fit component temperatures, radii, and extinction. We began by calculating synthetic contrasts and unresolved photometry, then downsampled the model spectrum to the instrumental resolution before calculating the composite \cs\ by comparing the synthetic data to the data set comprised of the spectrum and the two photometric components. We weighted the \cs\ contributions of the combined photometric data set and the single spectrum equally, to avoid bias in the \cs\ calculation that could result from the spectrum having many apparent degrees of freedom (2048 pixels) but very few true free parameters (\teff, surface gravity, and metallicity, to first order).

We found the initial best-fit parameters using a modified Gibbs sampler, which is a common MCMC optimization algorithm that we modified such that it could only move to lower \cs\ values instead of occasionally preferring a guess with a higher \cs\ value. We typically sampled with 150 walkers initialized with a random uniform distribution across the permitted parameter space (3000 $<$ \teff $<$ 7000 K; $0.05 < R < 2$\rsun, $0.1 \leq A_{V} \leq 0.5$), and ran the optimization until the fitter reached 400 steps without additional improvement in the \cs\ value. After optimizing, we used \emcee\ \citep{Foreman-Mackey2013} initialized with the 30\% of optimized walkers with the lowest \cs\ and ran it for 15,000 steps or until it reached convergence as determined by an autocorrelation time criterion, discarding the first 350 steps as burn-in, to assess the statistical error in our measurement and retrieve a final best-fit set of temperatures, radii, and extinction. During each fit we normalized the continuum of the data to match the continuum of the model using a low-order polynomial fit to the data, to compensate for any changes in the continuum slope from instrumental error.

In \citet{Sullivan2022b} we did not impose any evolutionary model-based priors on the temperature and radius retrieval because we were validating our method and could use unconstrained fits (i.e., those run with uniform priors on all parameters) as an independent check of the method. When performing the fits for that work we found that we typically recovered radius ratios that were greater than one, which was inconsistent with our expectations that most systems would be comprised of coeval main-sequence stars, where the secondary should be smaller and cooler than the primary. We concluded that this discrepancy was a result of inconsistency between the optical contrasts (typically $\Delta$optical $>> 0$) and the NIR contrasts (typically $\Delta$ NIR $\sim 0$), causing secondary stars to appear cooler and larger than their primaries. 

In the current work, we needed to recover the correct radius ratios to accurately calculate the corrected planetary radius. Therefore, we imposed a prior on the radius and radius ratio measurements. We assumed that the systems all had an age of 1 Gyr, and used the appropriate isochrone from the MIST stellar evolutionary models \citep{Paxton2011, Paxton2013, Paxton2015, Dotter2016, Choi2016}. For each best-fit temperature we calculated the predicted model radius and radius ratio, and imposed a Gaussian prior on the fitted stellar radii with a mean of the predicted model results and a standard deviation of 5\%, which is comparable to the accuracy of the best-available radius measurements \citep[e.g.,][]{Mann2015}.

To retrieve accurate parameters, especially for the high-\teff\ systems, which were at larger distances than the cooler stars, we had to fit for extinction, which \citet{Sullivan2022b} found required an informed prior to be constrained. We implemented a prior in A$_{V}$ using the system distances and the 3D dust map from \citet{Green2019} implemented in the \texttt{dustmaps} package\footnote{\url{https://dustmaps.readthedocs.io/en/latest/index.html}} \citep{Green2018}. We imposed a Gaussian prior using the mean and standard deviation of the samples at the appropriate location in 3D space as the parameters for the Gaussian prior. Typical values for the mean E$(g-r)$ were 0.05 mag, with a standard deviation of 0.02 mag. We converted from the \texttt{bayestar} units of $E(g-r)$ to units of \av\ using the equations in \citep{Green2018b}\footnote{\url{argonaut.skymaps.info/usage}}, which found that $E(B-V) = 0.884*$[Bayestar2019]. We converted to \av\ assuming $R_{V}$ = 3.1. 

\section{Results}
\begin{deluxetable*}{CCCCCCCCC}
\tablecaption{Stellar parameter fit results for all Earth analog hosting binary KOIs}
\tablecolumns{9}
\tablewidth{0pt}
\tablehead{
\colhead{KOI} & \colhead{T1} & \colhead{T2} & \colhead{T$_{Kepler}$} & \colhead{R1} & \colhead{R2/R1} & \colhead{R$_{Kepler}$} & \colhead{f$_{corr, p}$} & \colhead{f$_{corr, s}$}\\
\colhead{} & \colhead{(K)} & \colhead{(K)} & \colhead{(K)} & \colhead{\rsun} & \colhead{} & \colhead{\rsun} & \colhead{} & \colhead{}
}
\startdata
\input{star_params.txt}
\enddata
\tablecomments{An asterisk (*) denotes a system where a Gaia parallax was available.}
\label{tab:star results}
\end{deluxetable*}

\begin{figure}
    \plotone{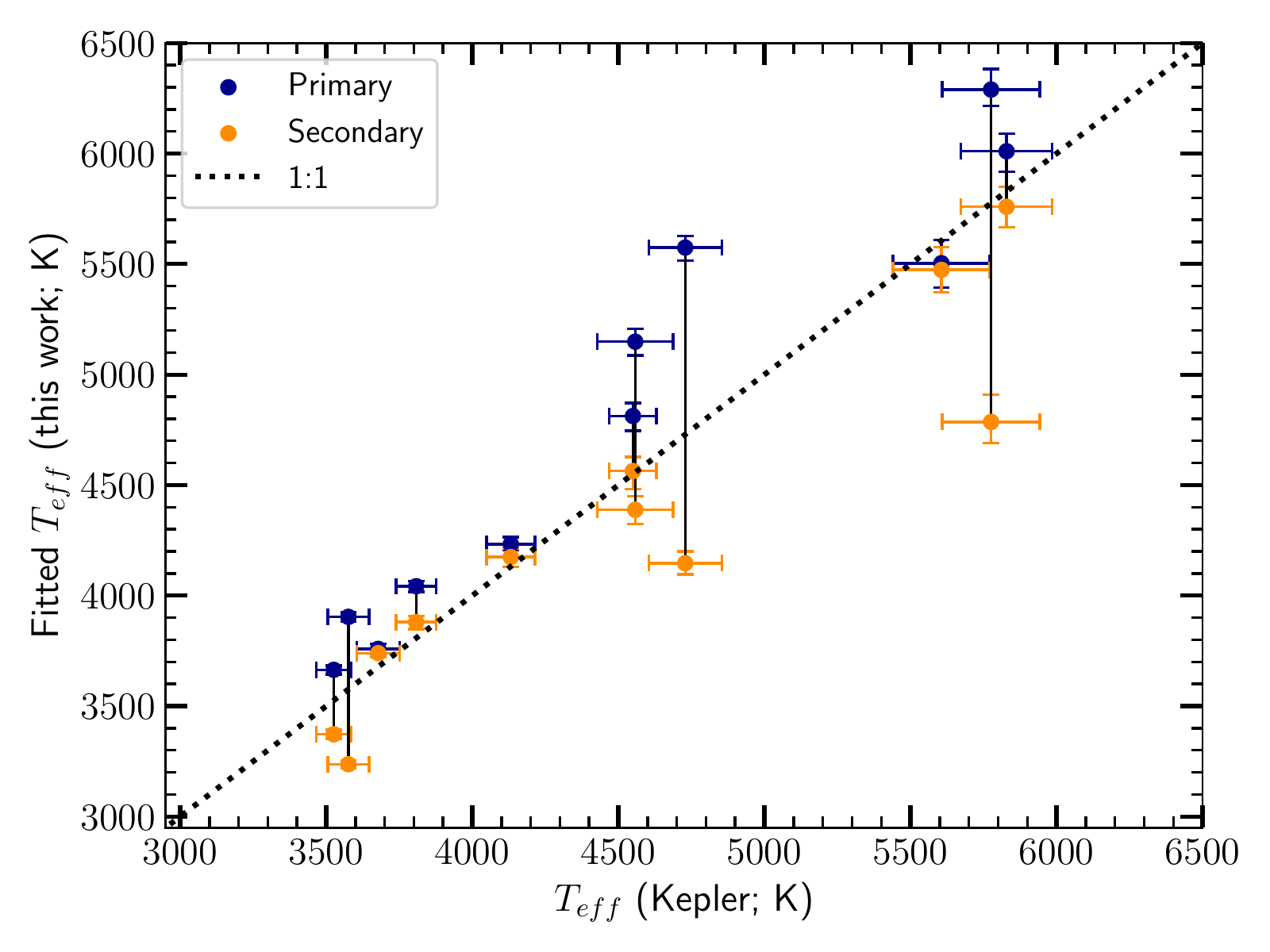}
    \caption{The temperature for the primary (blue) and secondary (orange) components of each system plotted against the \kep\ temperature. The solid black lines connect the components of each system, while the dotted diagonal black line indicates the 1:1 correspondence line. The temperatures typically change so that the primary component is hotter than the \kep\ temperature, and most of the changes are larger than the error bar on the measurement.}
    \label{fig:teff diff}
\end{figure}

To assess the inclusion of 17 near-Earth analog planets in the \ee\ calculation, we developed and implemented a Bayesian MCMC sampler to retrieve the individual temperatures and radii for close, spectroscopically unresolved binary stars hosting at least one small planet in or near the HZ ($R < 1.8 R_{\earth}; S < 5 S_{\earth}$). We observed 11 such systems using LRS2-R on the Hobby-Eberly Telescope, and analyzed them using those spectra, unresolved catalog photometry, and resolved component contrasts from archival NIR AO and optical speckle imaging. The following section presents our results for both the revised stellar parameters and the consequent revised planetary parameters.

Table \ref{tab:star results} summarizes the revised component stellar properties, and Appendix \ref{sec:appendix} shows summary plots for all of our fits as Figure Sets in the online version of this article. On average, the primary star temperatures were revised above the unresolved temperature by a median of +234$^{+310}_{-141}$ K and the secondary star temperatures were revised below the unresolved temperature by a median of -132$^{+39}_{-412}$ K. Figure \ref{fig:teff diff} shows a visual comparison of the component temperatures versus the \kep\ measured temperature for all our stellar systems. The large RMS spread in the average measurements are a result of a few high-contrast systems where both the primary and secondary star temperatures were significantly different from the \kep\ temperatures. 

We calculated the analytic planetary radius correction factor $f_{corr}$ for the cases where the primary or secondary were assumed to be the host star, where $R_{p, true} = f_{corr}R_{p, obs}$. If the primary star is the planet host, the correction factor is $f_{corr, pri} = (\frac{R_{pri}}{R_{kep}})\sqrt{1 + 10^{-0.4\Delta m}}$ \citep{Ciardi2015}. This differs slightly from the equation presented in \citet{Furlan2017}, because we assumed that the stellar radius measured when taking multiplicity into account was significantly different from the \kep\ measured radius. The correction factor if the secondary star is the planet host is $f_{corr, sec} = (\frac{R_{sec}}{R_{kep}})\sqrt{1 + 10^{+0.4 \Delta m}}$ \citep{Ciardi2015, Furlan2017}. The planetary radii were revised upward by an average factor of 1.43$^{+0.22}_{-0.21}$ if the primary star is the planet host and upward by an average factor of 1.69$^{+1.13}_{-0.15}$ if the secondary star is the planet host.

\begin{deluxetable*}{CCCCCCCCCC}
\tablecaption{Planet parameter fit results for all Earth analog hosting binary KOIs}\label{tab:planet_results}
\tablecolumns{10}
\tablewidth{0pt}
\tablehead{
\colhead{KOI} & \colhead{R$_{p, pri}$} & \colhead{R$_{p, sec}$} & \colhead{R$_{Kep}$} & \colhead{T$_{eq,pri}$} & \colhead{T$_{eq,sec}$} & \colhead{T$_{eq, Kep}$} & \colhead{S$_{pri}$} & \colhead{S$_{sec}$} & \colhead{S$_{kep}$}\\
\colhead{} & \colhead{($R_{\earth}$)} & \colhead{($R_{\earth}$)} & \colhead{($R_{\earth}$)} & \colhead{(K)} & \colhead{(K)} & \colhead{(K)} & \colhead{(S$_{\earth}$)} & \colhead{(S$_{\earth}$)} & \colhead{(S$_{\earth}$)}
}
\startdata
\input{revised_radii_teq.txt}
\enddata
\end{deluxetable*}

Using the revised stellar temperatures and radii, we calculated the revised instellation flux of the planets for the cases where the primary or secondary star was the planet host. We used the orbital period from ExoFOP and the best fit stellar mass calculated from 2 Gyr MIST models using the revised stellar temperature to calculate the corrected semimajor axis of each planet. The primary star masses were revised upward by (12$^{+26}_{-2}$)\% on average, while the secondary star masses were revised downward by(-2.5$^{+11}_{-11}$)\% on average. We calculated a best-fit luminosity using the 2 Gyr MIST models using the revised stellar temperature, and used the new luminosity and semimajor axis to calculate the revised instellation as 
\begin{equation*}
    \frac{S}{S_{\earth}} = (\frac{L}{L_{\sun}})(\frac{AU}{a})^{2}.
\end{equation*}
If the primary star is the planet host in all cases, the instellation flux was revised upward by an average of $32^{+178}_{-10}$\% relative to the \kep\ measured instellation. If the secondary star is the planet host in all cases, the instellation flux was revised downward by $-32^{+12}_{-307}$\% relative to the \kep\ measured instellation. This is likely a combination of the revised stellar parameters and the new \textit{Gaia} distances for most of our systems.

Table \ref{tab:planet_results} lists the radii, equilibrium temperatures, and instellation fluxes for all confirmed planets and planet candidates in our sample, including the \kep\ radius, instellation (S$_{\earth}$), and equilibrium temperature (T$_{eq}$); and the revised radius, instellation, and T$_{eq}$ if the primary or secondary star is the planet host. In each system, at least one planet fit our original selection criteria ($R < 1.8 R_{\earth}$, $S < 5 S_{\earth}$) before parameter revision, but all planets in multiple-planet systems are included in the analysis regardless of whether they fit the selection criteria. We calculated the errors on each parameter by assuming that the \kep\ measured values were described by a Gaussian posterior distribution with a mean and standard deviation equal to the published \kep\ value and its error, respectively. We calculated each new revised parameter using the relevant posteriors from our calculations and the presumed \kep\ posteriors, then took the 50th, 16th, and 84th percentile of the final posterior to be the most probable value and its lower and upper errors, respectively. 

\begin{figure}
    \gridline{\fig{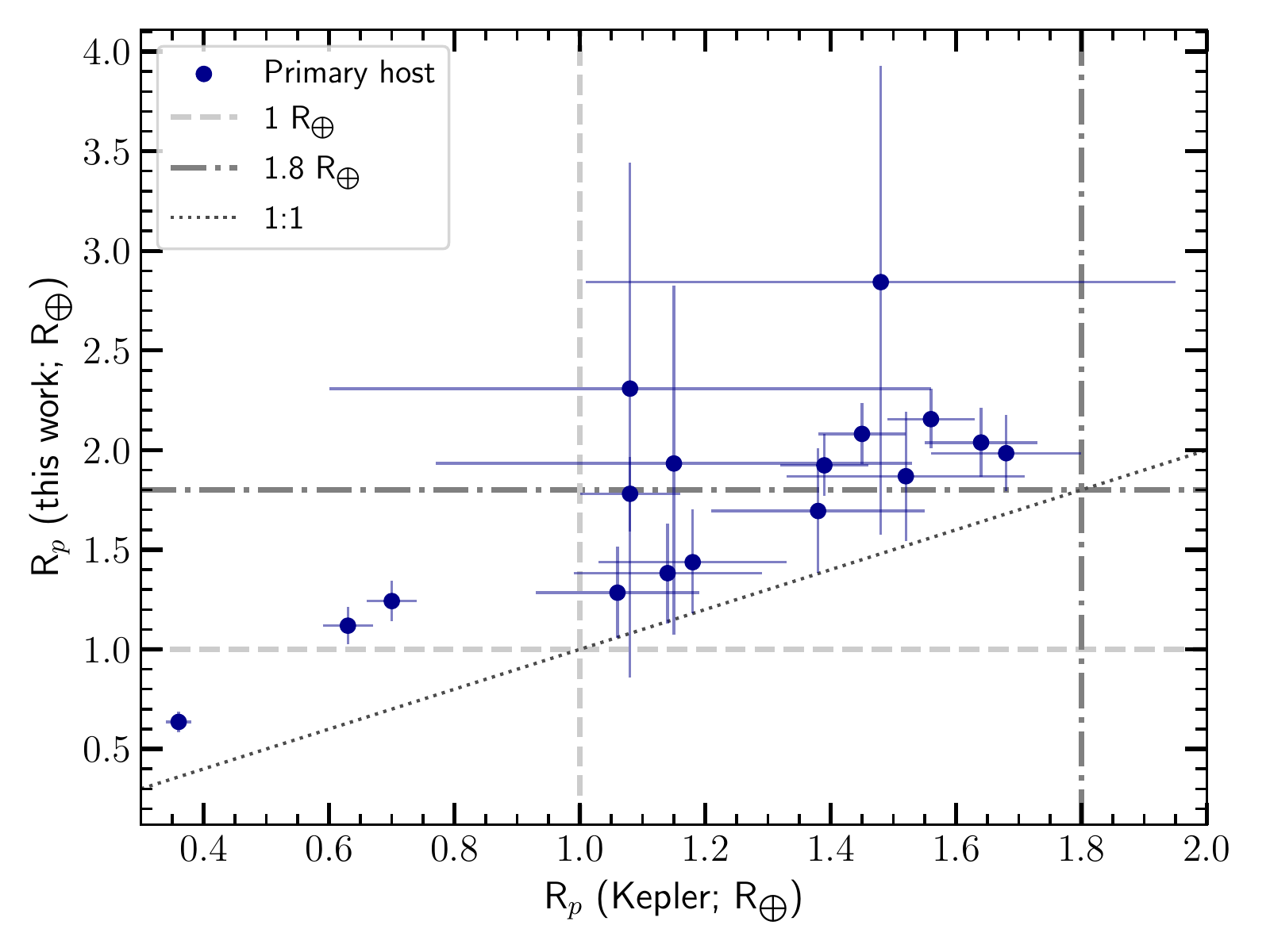}{0.95\linewidth}{}
	}
    \gridline{\fig{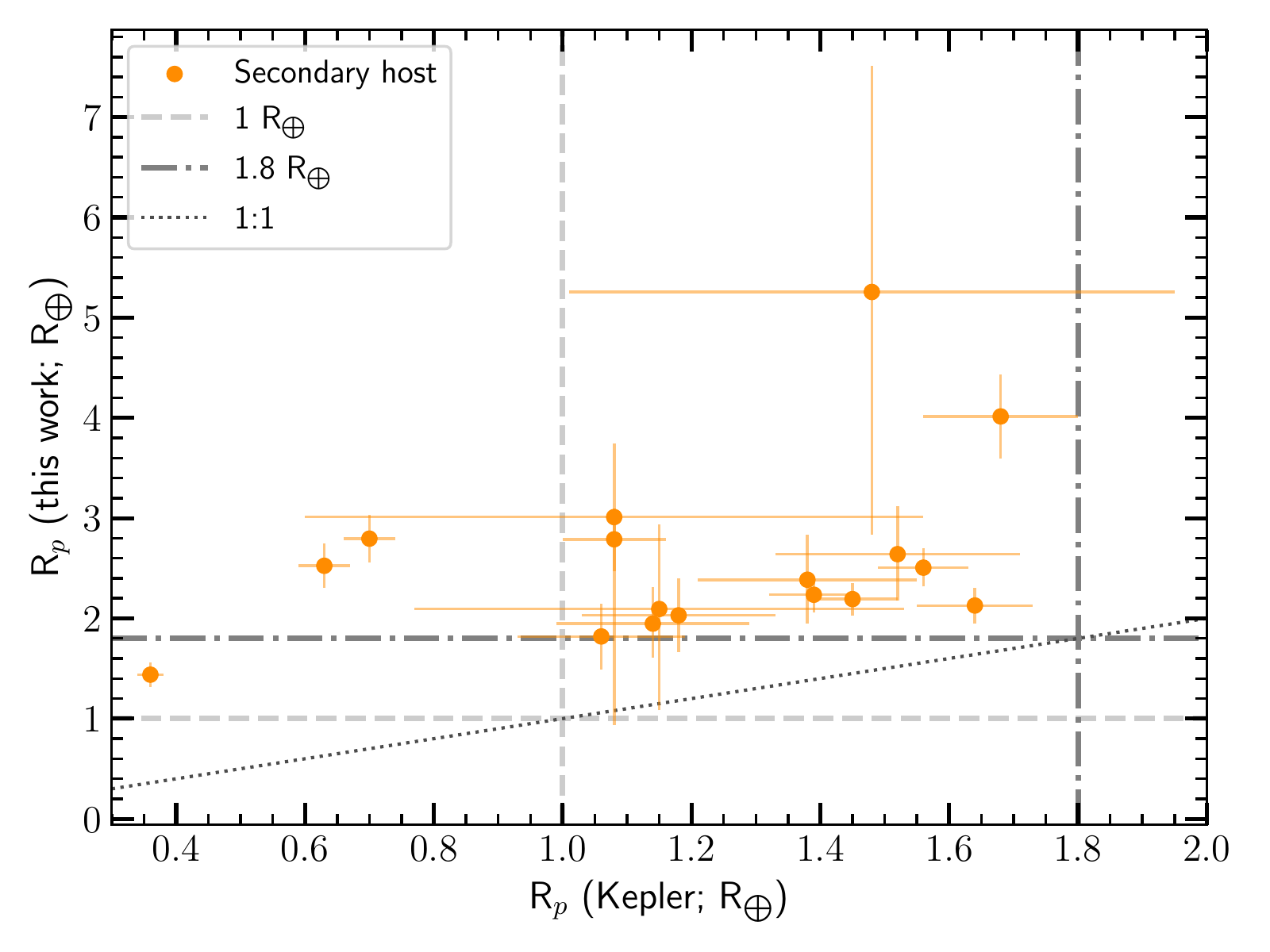}{0.95\linewidth}{}
}
    \caption{Top: The planetary radii measured from this work assuming the primary star is the planet host, plotted against the \kep\ measured radii. The revised radii are systematically larger than the \kep\ radii, with many revised radii landing in or above the radius gap. at 1.8R$_{\earth}$. The gray lines denote 1 R$_{\earth}$ (light gray), 1.8R$_{\earth}$ (dark gray), and the 1:1 correspondence line that indicates identical recovery for the two methods (black). Bottom: The same figure but if the secondary stars are the planet hosts. Many of the planets would fall above the radius gap at 1.8 R$_{\earth}$ if the secondary was their host star.}
    \label{fig:rp_results}
\end{figure}

Figure \ref{fig:rp_results} shows the revised planetary radii from our analysis plotted against the \kep\ measured radii for the cases where the primary and secondary stars are the planet hosts. The revised planet radii fall above the 1:1 correspondence line in all cases, and many systems in both the primary and secondary host cases fall above the radius gap at $\sim 1.8 R_{\earth}$ \citep[e.g.,][]{Petigura2013b, Fulton2017}, indicating that they likely have substantial hydrogen/helium atmospheres and are not suitable analogs for rocky planets. If the planets orbit the primary stars, $\sim$59\% (10/17) move into or above the radius gap, while if the planets orbit the secondary stars $\sim$ 94\% (16/17) move into or above the radius gap.

\section{Discussion and Conclusions}
We have retrieved new stellar parameters for 11 \kep\ binary-star systems that host at least one planet falling in or near the HZ ($R_{p} < 1.8R_{\earth}$; $S_{p} < 5 S_{\earth}$). Using the revised stellar temperatures and radii, we have revised 17 planets' radii and instellation flux, considering both possible scenarios for which component of the stellar binary is the planet host. 

\subsection{Revising the Properties of a Sample of Near-Earth Analog Planets in Binary Systems}
Small planets in or near the HZ are rare; in a search of the \citet{Berger2020} catalog of revised \kep\ planetary radii, there are 56 planets with $R_{p} < 1.8 R_{\earth}$ that were classified as being in the HZ. \citet{Ware2022} estimated that approximately 30 known Earth-analog planets fall in the HZ. There has been considerable effort exerted to calculate \ee\ using the small sample of rocky HZ planets (Earth analogs hereafter), but measurements of \ee\ have large errors (lack precision) and can substantially disagree between different works (lack accuracy). The cause of the lack of precision is the small sample of Earth analogs \citep[e.g.,][]{Dressing2013, Gaidos2013, Bryson2021, Ware2022}, while the lack of agreement between different estimates is based in the various completeness corrections that different groups impose and the different samples and planet parameters they have used \citep[e.g.,][]{Catanzarite2011, Petigura2013, Foreman-Mackey2014, Dressing2015, Burke2015, Silburt2015, Kunimoto2020, Bryson2021}. 

Although the subsample of near-Earth analogs in binary systems is relatively small, the addition or removal of even a small number of planets from the \ee\ calculation is important. Out of the 17 planets in our sample, we found that 10 of them fall in or above the radius gap even if the primary star is the planet host (Figure \ref{fig:rp_results}), and 16 out of 17 fall in or above the radius gap if the secondary stars are the planet hosts. As of 20220421, there are 154 \kep\ planets that fulfill our original sample selection criteria ($R_{p} \leq 1.8 R_{\earth}; S_{p} \leq 5 S_{\earth}$; \citealt{exoplanetarchive}), meaning that at least 10\% of the total sample of near-Earth analogs was removed because of our parameter revision, although that does not necessarily propagate to a 10\% revision of the \ee\ calculation, since some binaries were already excluded in \ee\ samples. Because of the high frequency of binaries ($\sim 50$\% for Sun-like stars; \citealt{Raghavan2010}) it is likely that there are more unrecognized binaries in the eta-Earth sample that must be identified and either re-characterized or removed.

\begin{figure*}
    \plottwo{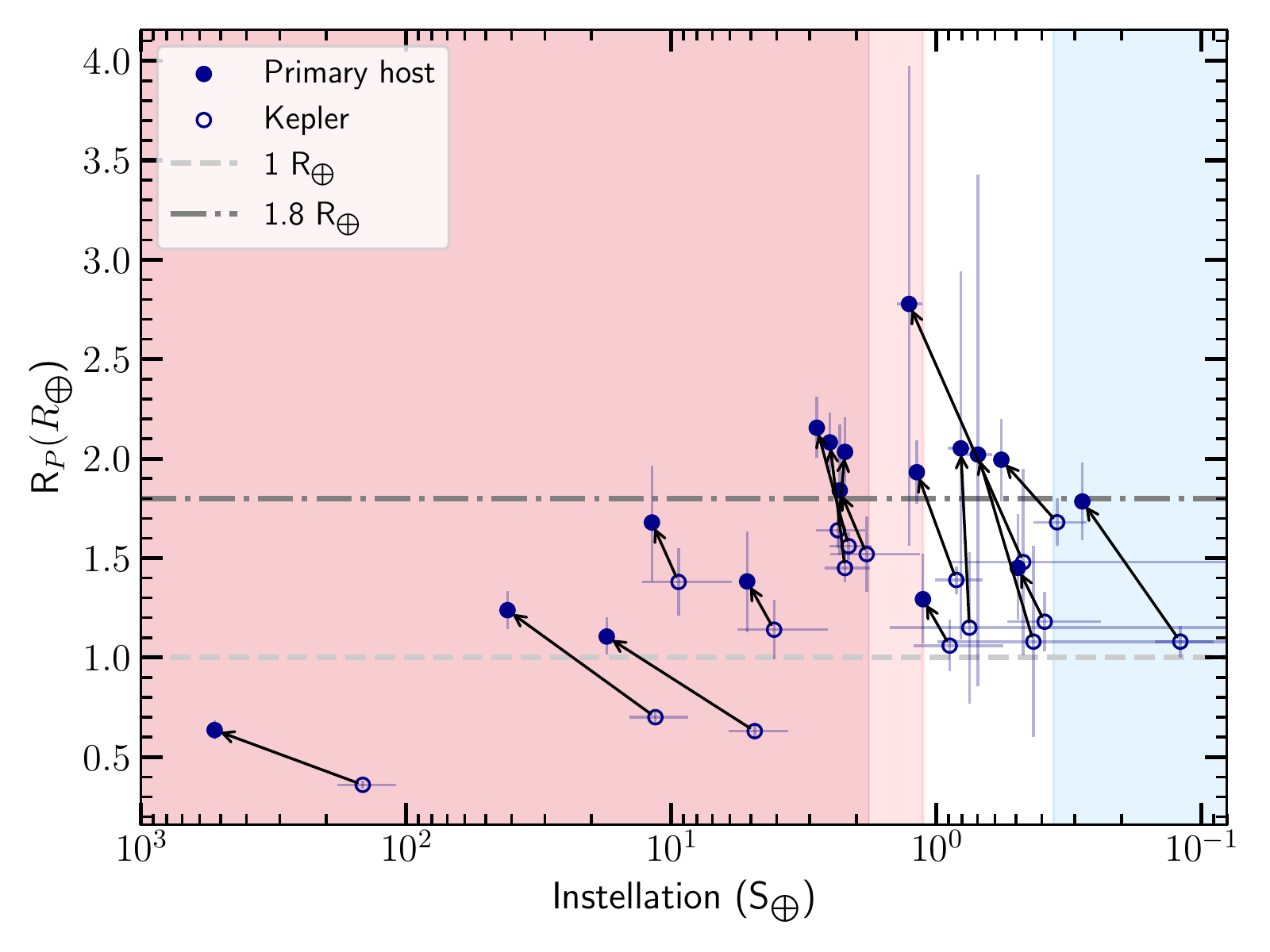}{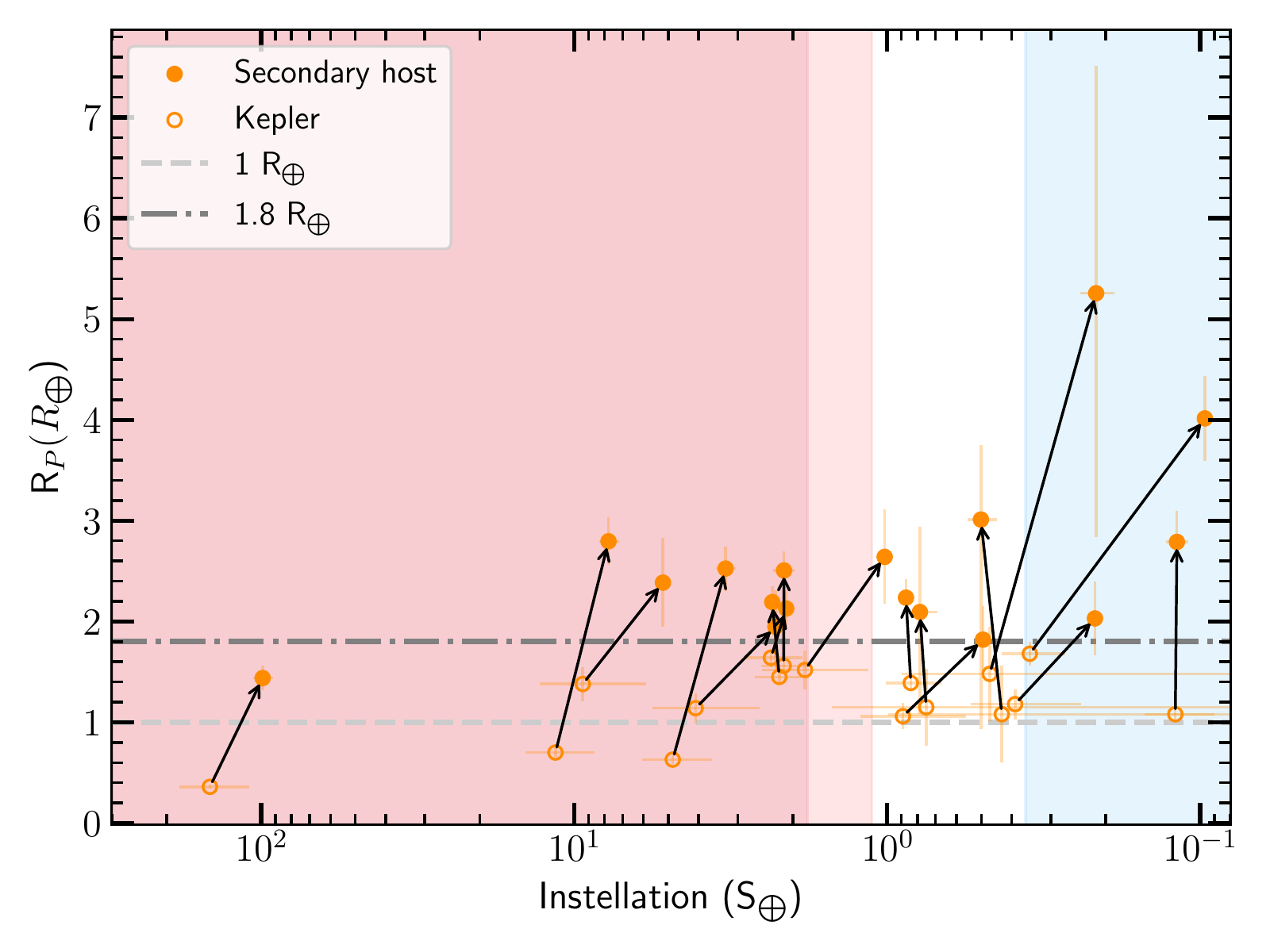}
    \caption{Left: Planetary radius if the primary star is the planet host, plotted against the instellation flux for the planet. The open circles denote the \kep\ values, while the closed circles denote the revised values from this work. The dark red shaded region is the region inside the optimistic HZ, while the light red shaded region is the instellation range between the conservative and optimistic HZ. The blue shaded region falls outside both the conservative and optimistic HZs. The light and dark gray dotted lines show the 1 and 1.8 R$_{\earth}$ boundaries, respectively. One of the planets originally outside the HZ moves into it. Almost all the planets in the HZ move above the radius gap. Right: The same figure but if the secondary star is the planet host. All but one of the planets fall above the radius gap if the secondary star is the planet host, and no Earth-size planets remain in the HZ.}
    \label{fig:rp_vs_hz}
\end{figure*}

Figure \ref{fig:rp_vs_hz} shows the planetary radii and instellation for the planet sample, as well as plotting the conservative (runaway greenhouse limit) and optimistic (recent Venus limit) inner edge of the HZ and the outer edge (maximum greenhouse limit) of the HZ \citep{Kopparapu2014}. Using the \kep\ parameters for the planets, 6 of the 17 fell into the conservative HZ as defined by \citet{Kopparapu2014}, and were near-Earth size ($R_{p} < 1.5 R_{\earth}$). After our parameter revision, 5 planets fell into the either the conservative or optimistic HZ but above the radius gap, and 2 planets fell into the conservative HZ and remained below the radius gap, but still had revised radii that were larger than the \kep\ values. Because the primary star temperature is typically higher than the \kep\ composite measured temperature, the majority of planets are more irradiated if the primary star is the planet host than they would have been around a single star with the \kep\ parameters. If the secondary star is the planet host almost all the planets in the sample fall above the radius gap and move toward smaller instellation fluxes, because the secondary star temperature is typically lower than the \kep\ measured composite temperature.

One of the purposes of this work was to explore how many planets should be removed from the \ee\ calculation after our re-analysis of their properties, but another purpose was to identify planets that could still be included in \ee\ even though they are in binary star systems. As high-resolution imaging follow-up of \kep\ targets has proliferated, many occurrence-rate calculations have dealt with planets in binaries by simply removing them from the sample under the premise that they will bias the resulting analysis. In regions of parameter space that are densely populated with planets this may be a feasible approach, but in the low-completeness regime of (near-)Earth analogs, every possible planet should be included in calculations to enhance the statistical power of the analysis. 

Therefore, we note that we found 2 planets or planet candidates (KOIs 1422.05 and 7235.01) that continued to have near-Earth radii and appeared to exist in the HZ of their host star if the primary star is the planet host. Future rocky-planet occurrence rate calculations should consider including these systems, but assessing whether they truly fall in the HZ (and therefore should be considered good near-Earth analogs) is complicated, and outside the scope of this paper. We assessed the HZ for each binary assuming that the radiation from the other star in the binary was negligible, which is appropriate if the primary star is the planet host, since the flux contribution from the secondary star is typically small if the planet is in the dynamically stable regime \citep[e.g.,][]{Simonetti2020}, which should be expected for main-sequence systems where systems have survived on $\sim$ Gyr timescales. However, if any of the planets are around the secondary stars, a more detailed analysis of the HZ location for those systems would be necessary. Even if the planets are around the primary star, the secondary star might also have more X-ray/UV flux than the primary (e.g., if the secondary is an active M dwarf), making its effect on potential habitability of the planet non-negligible even if it is much fainter than the primary in the visible.

\subsection{Conclusions} \label{sec:conclusions}
We used simultaneous fitting of unresolved low-resolution spectroscopy from the HET, catalog unresolved photometry, and archival NIR AO and optical contrasts to analyze 11 binary KOIs hosting near-Earth analog planets in or near the HZ. We retrieved the temperatures and radii of the components of each binary star, and used the revised parameters along with MIST stellar evolutionary models to revise the radii and instellation fluxes of the planets in each system. For each planet, we assessed whether it was moved in or out of the HZ, and whether it moved above the radius gap and so was no longer a rocky planet candidate.

We found that more than half of the planets in our sample had revised radii that were larger than 1.8 R$_{\earth}$, including 4 of the planets that were initially in the HZ, were in the HZ after parameter revision, or both. For most planets, the radiation environment did not change significantly, but the planetary radius did. This indicates that the first-order revisions to the instellation flux did not substantially impact whether a planet was potentially habitable, but the planetary radius revision caused by altered stellar temperatures and radii did significantly change the planetary demographics. This was likely driven by the revised stellar radii, which were constrained by evolutionary models and spectroscopy, as opposed to less accurate radii achievable using the KIC photometry and pre-Gaia distance estimates \citep{Brown2011}.

Our results indicate that if planets around the primary star in binary systems appear to be in the HZ, they likely are, but that they may not be rocky (and thus will not be Earth analogs that are suitable for inclusion in the \ee\ calculation). There is no systematic correction that can be applied to predict whether a supposedly rocky planet will have revised parameters that are suggestive of a substantial atmosphere. The revised planet parameters are dependent on both the revised stellar temperatures and radii, meaning that a full joint re-analysis of each system must be performed to correctly retrieve the stellar properties. However, near-Earth analog planets in binaries can contribute to the census of rocky exoplanets in HZs, making identification and vetting of such planets important. At a larger scale, binaries are extreme environments for planets to form in, so the sample of planets in binary stars is important to accurately characterize in its own right.\\ \\

K.S. acknowledges that this material is based upon work supported by the National Science Foundation Graduate Research Fellowship under Grant No. DGE-1610403. The authors thank the observing staff and resident astronomers at the Hobby-Eberly Telescope for obtaining the observations presented in this work. We especially thank Danny Krolikowski, Steven Janowiecki and Greg Zeimann for their assistance with the telluric corrections. The authors acknowledge the Texas Advanced Computing Center (TACC) at The University of Texas at Austin for providing high-performance computing resources that have contributed to the research results reported within this paper. 

The Hobby-Eberly Telescope (HET) is a joint project of the University of Texas at Austin, the Pennsylvania State University, Ludwig-Maximilians-Universität München, and Georg-August-Universität Göttingen. The HET is named in honor of its principal benefactors, William P. Hobby and Robert E. Eberly. The Low Resolution Spectrograph 2 (LRS2) was developed and funded by the University of Texas at Austin McDonald Observatory and Department of Astronomy and by Pennsylvania State University. We thank the Leibniz-Institut für Astrophysik Potsdam (AIP) and the Institut für Astrophysik Göttingen (IAG) for their contributions to the construction of the integral field units. 

This publication makes use of data products from the Two Micron All Sky Survey, which is a joint project of the University of Massachusetts and the Infrared Processing and Analysis Center/California Institute of Technology, funded by the National Aeronautics and Space Administration and the National Science Foundation. This research has made use of the SVO Filter Profile Service (\url{http://svo2.cab.inta-csic.es/theory/fps/}) supported from the Spanish MINECO through grant AYA2017-84089. This research has made use of the VizieR catalogue access tool, CDS, Strasbourg, France (DOI : 10.26093/cds/vizier). The original description of the VizieR service was published in 2000, A\&AS 143, 23. This work has made use of data from the European Space Agency (ESA) mission {\it Gaia} (\url{https://www.cosmos.esa.int/gaia}), processed by the {\it Gaia} Data Processing and Analysis Consortium (DPAC, \url{https://www.cosmos.esa.int/web/gaia/dpac/consortium}). Funding for the DPAC has been provided by national institutions, in particular the institutions participating in the {\it Gaia} Multilateral Agreement. This research has made use of the Exoplanet Follow-up Observation Program website, which is operated by the California Institute of Technology, under contract with the National Aeronautics and Space Administration under the Exoplanet Exploration Program.

\software{
astropy \citep{astropy2013, astropy2018, astropy2022}, corner \citep{Foreman-Mackey2016}, emcee \citep{Foreman-Mackey2013}, matplotlib \citep{Hunter2007}, numpy \citep{Harris2020}, scipy \citep{Virtanen2020}
}

\appendix
\section{Diagnostic and Summary Plots for All Near-Earth Analog Systems} \label{sec:appendix}
\begin{figure*}
\gridline{\includegraphics[width = 0.45\linewidth]{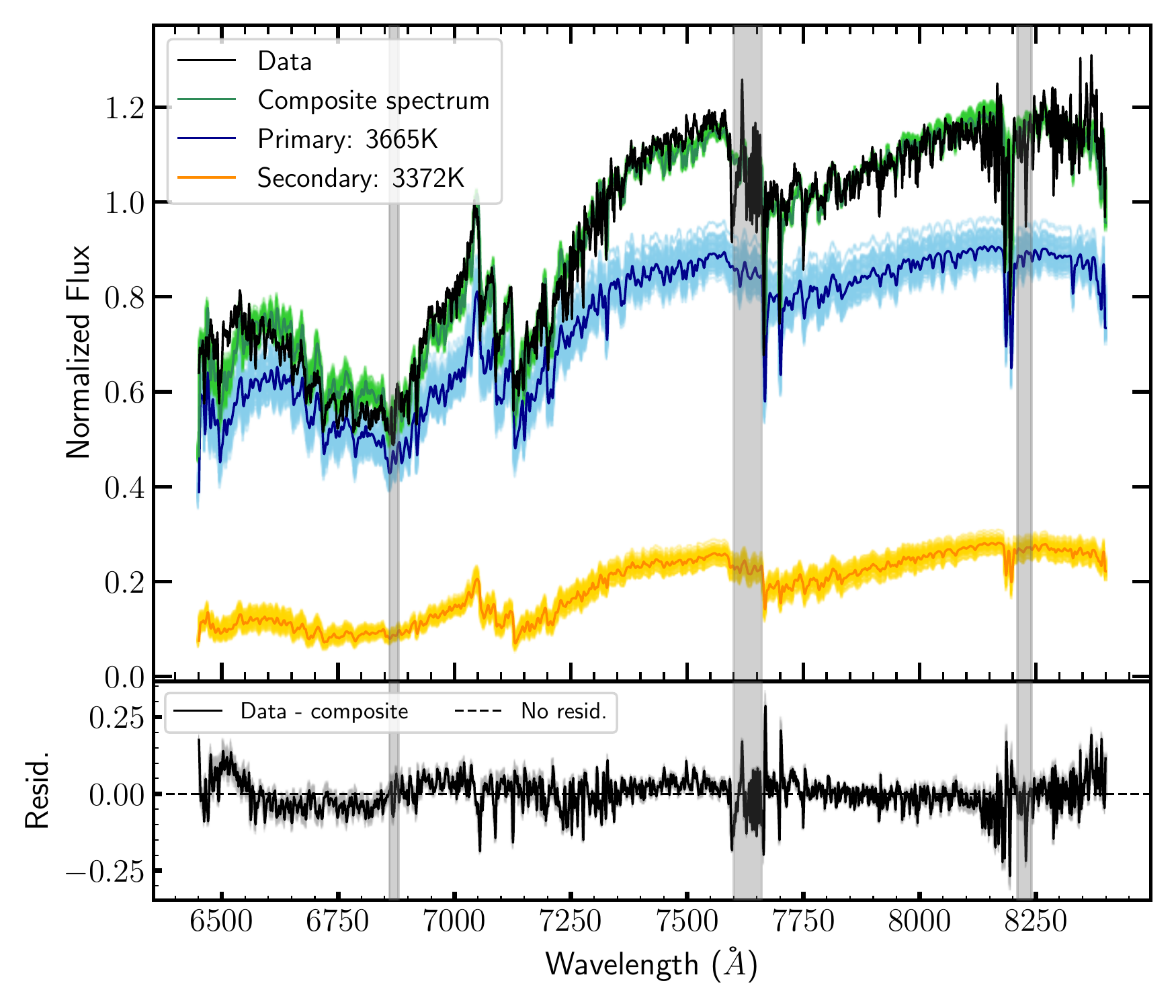} 
	\includegraphics[width = 0.45\linewidth]{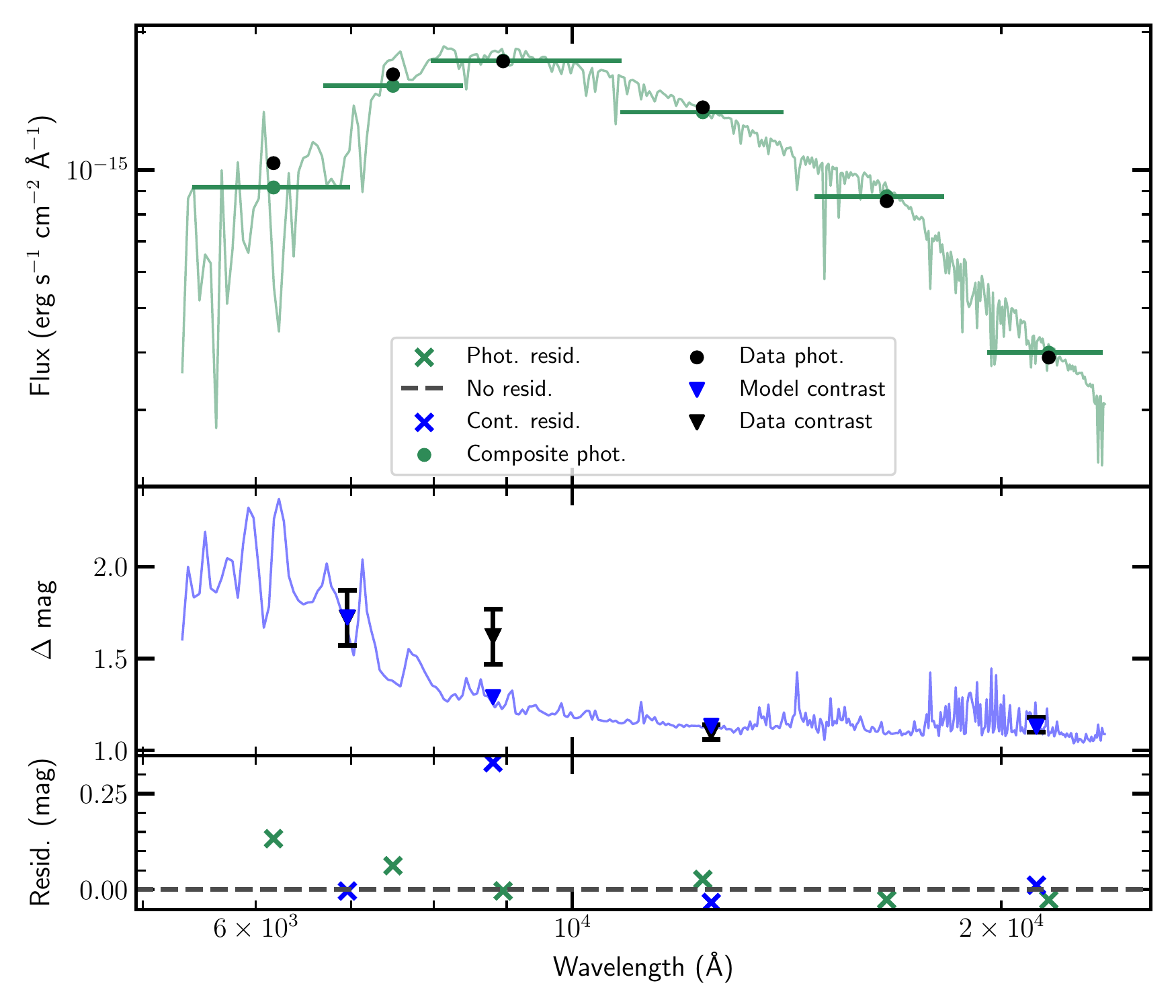}
}
\gridline{\includegraphics[width = 0.8\linewidth]{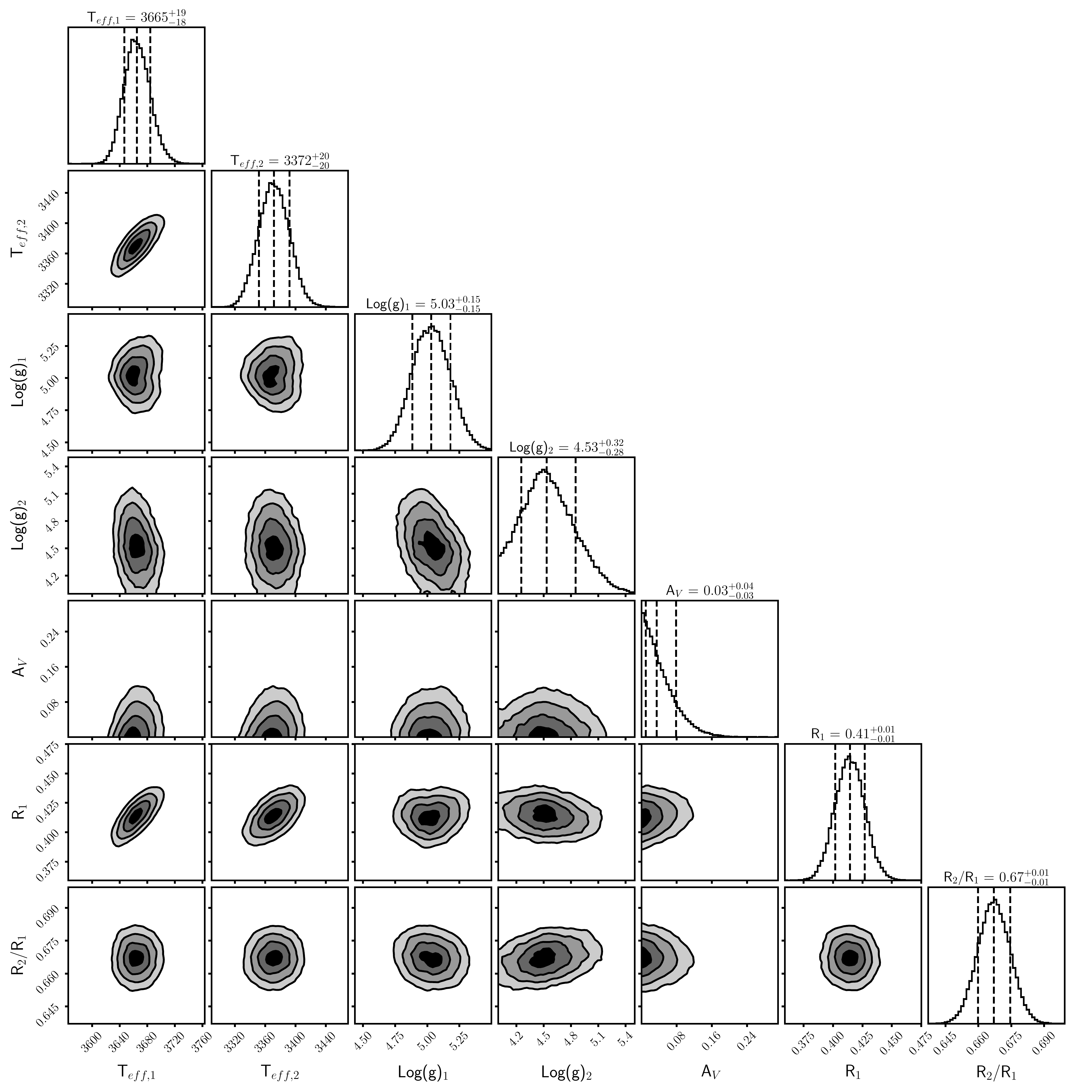}}
\caption{Summary and diagnostic plots for KOI-1422. The remainder of the systems have summary plots contained in a Figure Set in the online version of the article.}
\label{fig:1422 spec}
\end{figure*}

This appendix contains the diagnostic and fit summary plots for all the systems in our analysis. Figure \ref{fig:1422 spec} shows examples of the plots, and the remainder of the plots are shown in the Figure Set in the online version of the article. The figures resembling the top left panel of Figure \ref{fig:1422 spec} show the data spectrum, plotted with the best-fit composite spectrum and the best-fit component spectra. 100 random draws from the MCMC chains are plotted under each synthetic spectrum. The lower panel of the figure shows the residual from the best-fit synthetic composite spectrum (black) and 100 MCMC draws (gray) and the data. In general, the residual is dominated by a few poorly-modeled spectral features and noise, and there is no systematic slope that suggests that the (pseudo-)continuum is poorly fit. The gray bars running vertically through the figure indicate regions that were masked during the fitting to compensate for poorly-fit regions from the telluric correction. 

The figures resembling the top right panel of Figure \ref{fig:1422 spec} show the best-fit unresolved photometry (top section), contrasts (middle section) and residuals (bottom section) for each system. The green and blue underlaid lines are the best-fit composite spectrum and contrast curve, respectively. The black points are the data, and the green and blue markers are the best-fit values. The green bars in the top section denote the width of each photometric filter ($r'i'z'JHK_{s}$). In general the photometry and contrasts are well-fit, with residuals that are typically on the order of 0.1 mag or less. 

The figures resembling the bottom panel of Figure \ref{fig:1422 spec} show the marginalized posteriors (diagonals) and covariance between each pair of parameters for each system. The values above each column are the mean, 16th, and 84th percentiles, which we reported as the best-fit value and its error in Table \ref{tab:star results}. The posteriors are typically well-constrained and Gaussian, with the exception of surface gravity, which \citet{Sullivan2022b} found was not accurately retrieved in our fitting method, likely because of the low spectral resolution limiting the observable spectral features that are sensitive to surface gravity.

\bibliography{mcmc2_bib}
\end{document}

%% file: star_params.txt
1422 & 3664$^{+20}_{-18}$ & 3372$^{+20}_{-21}$ & 3526$\pm$60 & 0.41$^{+0.01}_{-0.01}$ & 0.67$^{+0.01}_{-0.01}$ & 0.38 $\pm$ 0.05 & 1.20$^{+0.18}_{-0.14}$ & 1.69$^{+0.25}_{-0.20}$\\
2124* & 4233$^{+28}_{-32}$ & 4174$^{+43}_{-40}$ & 4132$\pm$83 & 0.60$^{+0.01}_{-0.01}$ & 0.99$^{+0.01}_{-0.01}$ & 0.58 $\pm$ 0.03 & 1.43$^{+0.08}_{-0.08}$ & 1.51$^{+0.09}_{-0.08}$\\
2298* & 5575$^{+60}_{-53}$ & 4147$^{+51}_{-53}$ & 4729$\pm$125 & 0.89$^{+0.01}_{-0.01}$ & 0.68$^{+0.01}_{-0.01}$ & 0.52 $\pm$ 0.03 & 1.76$^{+0.10}_{-0.09}$ & 3.97$^{+0.25}_{-0.23}$\\
2418* & 3904$^{+22}_{-21}$ & 3237$^{+16}_{-15}$ & 3576$\pm$71 & 0.53$^{+0.01}_{-0.01}$ & 0.43$^{+0.01}_{-0.01}$ & 0.46 $\pm$ 0.03 & 1.17$^{+0.10}_{-0.07}$ & 2.37$^{+0.20}_{-0.17}$\\
2862 & 3759$^{+22}_{-23}$ & 3740$^{+19}_{-20}$ & 3678$\pm$73 & 0.46$^{+0.01}_{-0.01}$ & 0.98$^{+0.01}_{-0.01}$ & 0.51 $\pm$ 0.03 & 1.23$^{+0.09}_{-0.08}$ & 1.28$^{+0.09}_{-0.07}$\\
3010 & 4042$^{+25}_{-24}$ & 3880$^{+31}_{-27}$ & 3808$\pm$69 & 0.57$^{+0.01}_{-0.01}$ & 0.91$^{+0.02}_{-0.02}$ & 0.52 $\pm$ 0.03 & 1.39$^{+0.08}_{-0.08}$ & 1.60$^{+0.10}_{-0.09}$\\
3255 & 4812$^{+67}_{-58}$ & 4564$^{+82}_{-63}$ & 4550$\pm$81 & 0.72$^{+0.02}_{-0.02}$ & 0.97$^{+0.01}_{-0.01}$ & 0.68 $\pm$ 0.03 & 1.38$^{+0.08}_{-0.07}$ & 1.60$^{+0.10}_{-0.09}$\\
4986* & 6290$^{+73}_{-93}$ & 4786$^{+96}_{-123}$ & 5776$\pm$167 & 1.14$^{+0.04}_{-0.05}$ & 0.62$^{+0.01}_{-0.01}$ & 0.73 $\pm$ 0.23 & 1.66$^{+0.70}_{-0.41}$ & 3.19$^{+1.40}_{-0.78}$\\
5545* & 6011$^{+92}_{-80}$ & 5759$^{+93}_{-90}$ & 5829$\pm$156 & 1.01$^{+0.04}_{-0.04}$ & 0.89$^{+0.01}_{-0.01}$ & 0.80 $\pm$ 0.35 & 1.61$^{+1.26}_{-0.49}$ & 1.80$^{+1.37}_{-0.57}$\\
5971* & 5149$^{+62}_{-58}$ & 4388$^{+64}_{-61}$ & 4558$\pm$130 & 0.79$^{+0.01}_{-0.01}$ & 0.85$^{+0.01}_{-0.01}$ & 0.55 $\pm$ 0.04 & 1.65$^{+0.13}_{-0.11}$ & 2.56$^{+0.22}_{-0.19}$\\
7235* & 5504$^{+110}_{-104}$ & 5474$^{+102}_{-103}$ & 5606$\pm$166 & 0.84$^{+0.04}_{-0.03}$ & 0.98$^{+0.01}_{-0.01}$ & 0.76 $\pm$ 0.25 & 1.54$^{+0.86}_{-0.38}$ & 1.55$^{+0.72}_{-0.38}$\\

%% file: revised_radii_teq.txt
1422.01 & 1.69$^{+0.32}_{-0.31}$ & 2.39$^{+0.45}_{-0.44}$ & 1.38$\pm$0.17 & 484$^{+34}_{-34}$ & 364$^{+26}_{-25}$ & 446 & 11.82$^{+0.58}_{-0.58}$ & 5.20$^{+0.29}_{-0.28}$ & 9.38$\pm$3.49\\
1422.02 & 1.87$^{+0.32}_{-0.33}$ & 2.64$^{+0.48}_{-0.46}$ & 1.52$\pm$0.19 & 323$^{+21}_{-21}$ & 243$^{+16}_{-15}$ & 297 & 2.31$^{+0.11}_{-0.11}$ & 1.02$^{+0.06}_{-0.05}$ & 1.83$\pm$0.68\\
1422.03 & 1.38$^{+0.25}_{-0.25}$ & 1.95$^{+0.36}_{-0.34}$ & 1.14$\pm$0.15 & 395$^{+27}_{-26}$ & 297$^{+20}_{-20}$ & 363 & 5.17$^{+0.25}_{-0.25}$ & 2.27$^{+0.12}_{-0.12}$ & 4.09$\pm$1.53\\
1422.04 & 1.44$^{+0.26}_{-0.25}$ & 2.03$^{+0.37}_{-0.37}$ & 1.18$\pm$0.15 & 219$^{+15}_{-15}$ & 165$^{+11}_{-11}$ & 202 & 0.49$^{+0.02}_{-0.02}$ & 0.22$^{+0.01}_{-0.01}$ & 0.39$\pm$0.15\\
1422.05 & 1.28$^{+0.23}_{-0.23}$ & 1.82$^{+0.33}_{-0.33}$ & 1.06$\pm$0.13 & 269$^{+18}_{-18}$ & 202$^{+13}_{-14}$ & 248 & 1.12$^{+0.06}_{-0.05}$ & 0.49$^{+0.03}_{-0.03}$ & 0.89$\pm$0.33\\
2124.01 & 2.08$^{+0.16}_{-0.15}$ & 2.19$^{+0.16}_{-0.17}$ & 1.45$\pm$0.07 & 325$^{+11}_{-11}$ & 320$^{+11}_{-11}$ & 311 & 2.52$^{+0.10}_{-0.11}$ & 2.33$^{+0.14}_{-0.12}$ & 2.21$\pm$0.43\\
2298.01 & 1.24$^{+0.10}_{-0.10}$ & 2.80$^{+0.24}_{-0.24}$ & 0.70$\pm$0.04 & 720$^{+29}_{-28}$ & 442$^{+18}_{-17}$ & 469 & 41.43$^{+2.63}_{-2.64}$ & 7.76$^{+0.55}_{-0.57}$ & 11.47$\pm$2.87\\
2298.02 & 1.12$^{+0.09}_{-0.09}$ & 2.52$^{+0.22}_{-0.22}$ & 0.63$\pm$0.04 & 580$^{+23}_{-23}$ & 357$^{+15}_{-15}$ & 378 & 17.51$^{+1.11}_{-1.12}$ & 3.28$^{+0.23}_{-0.24}$ & 4.84$\pm$1.21\\
2298.03 & 0.64$^{+0.05}_{-0.05}$ & 1.44$^{+0.12}_{-0.12}$ & 0.36$\pm$0.02 & 1362$^{+53}_{-50}$ & 837$^{+33}_{-33}$ & 887 & 527.66$^{+33.53}_{-33.63}$ & 98.86$^{+7.03}_{-7.26}$ & 145.69$\pm$36.41\\
2418.01 & 1.98$^{+0.19}_{-0.19}$ & 4.01$^{+0.42}_{-0.42}$ & 1.68$\pm$0.12 & 230$^{+9}_{-10}$ & 125$^{+5}_{-5}$ & 196 & 0.57$^{+0.02}_{-0.03}$ & 0.10$^{+0.00}_{-0.00}$ & 0.35$\pm$0.08\\
2862.01 & 2.04$^{+0.17}_{-0.17}$ & 2.13$^{+0.18}_{-0.18}$ & 1.64$\pm$0.09 & 306$^{+12}_{-12}$ & 302$^{+12}_{-12}$ & 316 & 2.21$^{+0.12}_{-0.12}$ & 2.10$^{+0.10}_{-0.10}$ & 2.35$\pm$0.49\\
3010.01 & 1.92$^{+0.16}_{-0.15}$ & 2.24$^{+0.18}_{-0.17}$ & 1.39$\pm$0.07 & 270$^{+9}_{-9}$ & 248$^{+9}_{-9}$ & 244 & 1.18$^{+0.05}_{-0.05}$ & 0.87$^{+0.05}_{-0.05}$ & 0.84$\pm$0.17\\
3255.01 & 2.16$^{+0.15}_{-0.15}$ & 2.51$^{+0.19}_{-0.19}$ & 1.56$\pm$0.07 & 335$^{+11}_{-12}$ & 314$^{+12}_{-12}$ & 308 & 2.82$^{+0.19}_{-0.20}$ & 2.13$^{+0.18}_{-0.16}$ & 2.14$\pm$0.38\\
4986.01 & 2.84$^{+1.09}_{-1.27}$ & 5.26$^{+2.26}_{-2.42}$ & 1.48$\pm$0.47 & 302$^{+41}_{-53}$ & 180$^{+25}_{-32}$ & 211 & 1.27$^{+0.14}_{-0.14}$ & 0.21$^{+0.03}_{-0.03}$ & 0.47$\pm$0.43\\
5545.01 & 2.31$^{+1.14}_{-1.45}$ & 3.01$^{+0.74}_{-2.08}$ & 1.08$\pm$0.48 & 262$^{+51}_{-65}$ & 237$^{+46}_{-59}$ & 206 & 0.70$^{+0.07}_{-0.08}$ & 0.50$^{+0.05}_{-0.06}$ & 0.43$\pm$0.56\\
5971.01 & 1.78$^{+0.19}_{-0.19}$ & 2.79$^{+0.31}_{-0.31}$ & 1.08$\pm$0.08 & 203$^{+10}_{-10}$ & 159$^{+8}_{-8}$ & 149 & 0.28$^{+0.02}_{-0.02}$ & 0.12$^{+0.01}_{-0.01}$ & 0.12$\pm$0.03\\
7235.01 & 1.93$^{+0.89}_{-0.86}$ & 2.09$^{+0.84}_{-1.01}$ & 1.15$\pm$0.38 & 258$^{+45}_{-48}$ & 255$^{+43}_{-49}$ & 237 & 0.81$^{+0.10}_{-0.10}$ & 0.79$^{+0.09}_{-0.10}$ & 0.75$\pm$0.75\\